\documentclass{aa}
\usepackage{graphicx}
\usepackage{txfonts}
\usepackage[citecolor=blue, linkcolor=blue, urlcolor = black, colorlinks = true]{hyperref}
\usepackage{ulem}
\usepackage{multirow}

\usepackage{graphicx}	% Including figure files
\usepackage{amsmath}	% Advanced maths commands
\usepackage{amssymb}	% Extra maths symbols
\usepackage{xspace}

\usepackage{siunitx}
\usepackage{comment}
\usepackage{textcomp}
\usepackage{bm}
\usepackage{lscape}
\usepackage{xcolor}
\newcommand{\eps}{\varepsilon}
\newcommand{\dt}[1]{\frac{\partial  #1}{\partial t}}
\newcommand{\vv}{\bold{v}}
\newcommand{\vn}{\vec{n}}

\newcommand{\Msun}{\,M$_{\odot}$\xspace}

\newcommand{\xcx}{$X_{\rm core}/X_{\rm ini}$\xspace}
\newcommand{\percent}{~per~cent\xspace}
\newcommand{\teff}{$T_{\rm eff}$\xspace}

\newcommand{\omegai}{$(\Omega_{\rm eq} / \Omega_{\rm c})_{\rm i}$\xspace}
\newcommand{\omegac}{$\Omega_{\rm eq} / \Omega_{\rm c}$\xspace}

\newcommand{\edit}[1]{#1}%{{\color{red} #1}}
\newcommand{\mesa}{\texttt{MESA}\xspace}
\newcommand{\ester}{\texttt{ESTER}\xspace}

\defcitealias{Burssens2023}{B23}

   \title{A two-dimensional perspective of the rotational evolution of rapidly rotating intermediate-mass stars}
   \subtitle{Implications for the formation of single Be stars}

   \author{J.S.G. Mombarg\inst{1}
   \and 
   M. Rieutord\inst{1}
   \and
   F. Espinosa Lara\inst{2}
          }
   \institute{IRAP, Universit\'e de Toulouse, CNRS, UPS, CNES, 14 avenue \'Edouard Belin, F-31400 Toulouse, France
   \and 
   Space Research Group, University of Alcal\'a, 28871 Alcal\'a de Henares, Spain\\
              \email{jmombarg@irap.omp.eu}
        }

   \date{Received 2 November 2023; accepted 15 January 2024}
\titlerunning{2D rotational evolution}
\authorrunning{Mombarg et al.}

\begin{document}  
  \abstract
   {Recently, the first successful attempt at computing stellar structure and evolution models in two dimensions has been presented with models that include the centrifugal deformation and self-consistently compute the velocity field.   }
   {This paper aims at studying the rotational evolution of two-dimensional models of stars rotating at a significant fraction of their critical angular velocity. From the predictions of these models, we aim to improve our understanding of the formation of single Be stars. }
   {Using the \ester code that solves the stellar structure of a rotating star in two dimensions with time evolution, we have computed evolution tracks of stars between 4 and 10\Msun for initial rotation rates ranging between 60 and 90\percent the critical rotation rate. Furthermore, we compute models for both a Galactic metallicity and an SMC metallicity.  }
   {A minimum initial rotation rate at the start of the main sequence is needed to spin up the star to critical rotation within its main sequence lifetime. This threshold depends on the stellar mass, and increases with increasing mass. The models do not predict any stars above 8\Msun to reach (near) critical rotation during the main sequence. Furthermore, we find the minimum threshold of initial angular velocity is lower for SMC metallicity compared to Galactic metallicity, which is in agreement with the increased fraction in the number of observed Be stars in lower metallicity environments. The strong difference in the rotational evolution between different masses is not predicted by any one-dimensional stellar evolution models.     }
   {Self-consistent two-dimensional stellar evolution provide more insight into the rotational evolution of intermediate-mass stars, and our predictions are consistent with observations of velocity distributions and fraction of Be stars amongst B-type stars. \edit{We find that stars with a mass above 8\Msun do not increase their fraction of critical rotation during the main sequence. Since a fraction of stars above 8\Msun have been observed to display the Be phenomenon, other processes or formation channels must be at play, or critical rotation is not required for the Be phenomenon above this mass.}}

   \keywords{stars: emission line, Be - stars: evolution - stars: interiors - stars: massive - stars: rotation 
               }

   \maketitle
%
%-------------------------------------------------------------------

\section{Introduction}
The rotational evolution of stars born with a convective core and radiative envelope is still a great unknown in the theory of stellar structure and evolution (SSE). Since a star's rotation profile (or shear profile rather) influences the efficiency of rotationally-induced chemical mixing \citep[e.g.][]{Zahn1992, Chaboyer1992}, uncertainties in the predicted rotation profile in turn introduces uncertainties on the evolutionary pathway of the star in the Hertzsprung-Russell diagram (HRD). Studies using rotating stellar models are limited to one (spatial) dimension and the validity of such models for stars that are spinning at a significant {fraction} of their critical rotation rate is questionable. In \cite{Mombarg2023b}, we have successfully ran the first two-dimensional SSE models of a 12\Msun star with the \ester code \citep{Espinosa2013, Rieutord2016}. We have calibrated our models with a $\beta$~Cephei pulsator that has been well-characterized by \cite{Burssens2023} and we have shown that all observationally derived quantities such as luminosity, effective temperature, core mass, and core- and surface rotation, could be reproduced by the \ester models. 

Rotation also plays an important role in the formation of classical Be stars \citep{Struve1931}, which we define here as stars of spectral type B that are rotating sufficiently fast to form a decretion disk. The formation of Be stars is not fully understood, and formation channels have been proposed including binary interaction \citep[e.g.][]{Kriz1975, Pols1991} or single star evolution \citep{Bodenheimer1995, Ekstrom2008, Granada2013, Hastings2020}. It is generally believed that the formation of a disk requires the star to spin at large fraction of its critical  {angular} velocity, but the exact limit is unknown and studies have argued that (near) critical rotation is not a necessity for the Be phenomenon based on measured subcritical rotation rates. While \cite{Townsend2004} argue that the observed subcritical rotation rates are underestimated because gravity darkening \citep{VonZeipel1924} has not been properly taken into account, later studies accounting for this effect still find subcritical rotation rates \citep[e.g.][]{Cranmer2005, Fremat2005}. Moreover, \cite{espinosa2011} have shown that the von Zeipel law with a 1/4-exponent, as used by \cite{Townsend2004}, overestimates the difference between the effective temperature at the pole and the equator, still reducing the impact of gravity darkening on V$\sin i$ measurements.

The present paper follows up the work of \cite{Mombarg2023b}. In this previous work, we discovered that a 12\Msun\ model never reaches a quasi-steady state, which would occur if nuclear evolution were extremely slow. It turns out that nuclear evolution of such a model is too fast, especially near the end of the main sequence, for the baroclinic waves to be damped. Indeed, as shown by \cite{busse81}, the initial distribution of angular momentum of a star relaxes to a steady state when all baroclinic waves are damped out. This damping occurs on a time scale\footnote{This time scale is sometimes related to the Eddington-Sweet time scale, but see the discussion of \cite{rieutord06}.} $t_{\rm relax}\sim (R^2/K)(N^2/\Omega^2)$, namely on the thermal relaxation time $R^2/K$ (where $R$ is the radius of the star and $K$ the heat diffusivity) times the square ratio of the Brunt-V\"ais\"al\"a frequency $N$ and the angular velocity $\Omega$. If the nuclear evolution is much longer than this relaxation time then the evolution of the star can be computed as a series of steady states with the same total angular momentum \cite[e.g.][]{Gagnier2019b}. In such a case, it has been shown that provided the initial angular velocity is fast enough, and mass-loss is negligible, the star inevitably reaches the critical rotation during the main sequence. The result of \cite{Mombarg2023b} nuances this simple picture showing that nuclear evolution can be fast enough to prevent an evolution in a series of relaxed quasi-steady states. The interesting conclusion is that even if initial angular velocity is high and mass-loss unimportant, the star may not reach its critical angular velocity. This is a crucial point to investigate if we want to understand the Be phenomenon. Hence, we shall here extend the investigations of \cite{Mombarg2023b} to the broad mass range of intermediate-mass stars, namely 4 to 10~\Msun, where nuclear evolution is slower and possibly longer than the relaxation time of baroclinic waves. The comparison of time scales is however not easy since parameters like $K$, $N$ or $\Omega$ vary by orders of magnitude inside a star. It turned out that the computation of the full evolution of 2D-ESTER models was the most direct way to get a broad view of the rotational evolution of these early-type stars, including the effects of metallicity.

Hence, we organised the paper as follows: in Section~\ref{sec:ester}, we briefly describe the physics of the \ester models that we use. In Section~\ref{sec:rot_evol_MW}, we present the predicted evolution of the rotation velocity at the stellar surface and compare these predictions with the measured projected surface velocities of B-type stars by \cite{Huang2010}. We discuss the effect of metallicity on the rotational evolution in Section~\ref{sec:rot_evol_SMC}, and finally present our conclusions in Section~\ref{sec:conclusions}.

%focuses on intermediate-mass stars between 4 and 10\Msun, thus covering a large fraction of the mass range of B-type stars. Here, we study single \sout{Be} \MR{B-type} stars which are thought be either born as fast rotators or have spun up during the main sequence phase as a result of efficient transport of angular momentum from the core towards the surface. We present the results of rotating 2D SSE models that account for the centrifugal deformation and self-consistently compute the velocity field from the baroclinic torque, something that is not possible with one-dimensional models. We aim to study the rotational evolution predicted by 2D SSE models and place these predictions in the context of observed characteristics of Be stars. Empirically derived upper limits on the fraction of the critical velocity, $v_{\rm eq}/v_{\rm c}$, for non-Be stars by \cite{Huang2010} suggests the threshold for the Be phenomenon is higher for lower-mass stars ($\lesssim 4\,{\rm M_\odot}$) compared to higher-mass stars ($\gtrsim 8.6\,{\rm M_\odot}$). Observed fractions of Be stars compared to the total number of B-type stars increases in environments with lower metallicity \citep{Maeder1999, Martayan2010, Iqbal2013}. In Section

\begin{figure}
    \centering
    \includegraphics[width = 0.9\columnwidth]{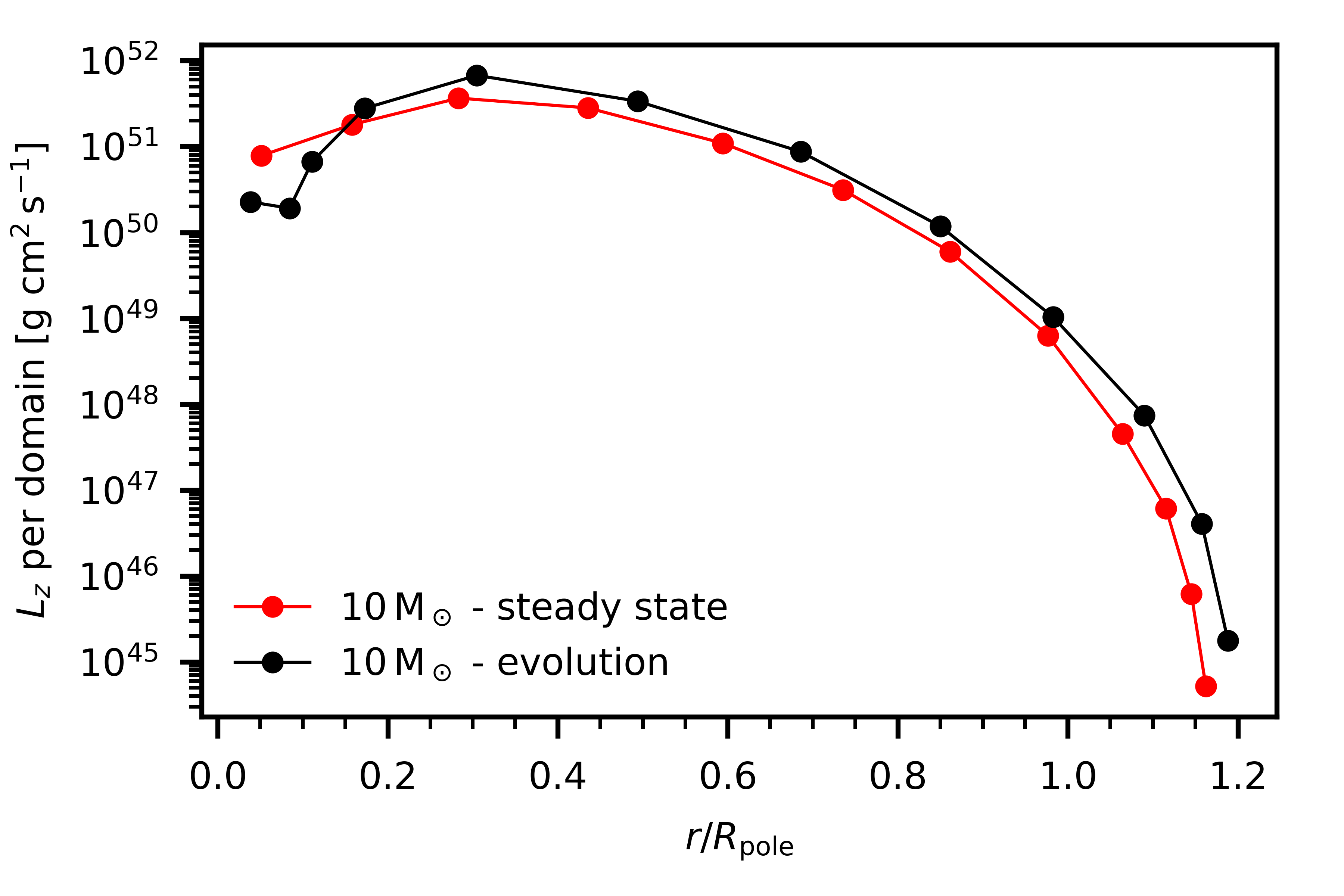}
    \caption{Angular momentum integrated per domain as a function of the radial coordinate in the equatorial direction. Once for a steady-state model (no viscous stresses), and once for the \ester models used in this work. Both models are computed at \xcx = 0.12 and have a rotation rate \omegac = 0.57. }
    \label{fig:AM_distr}
\end{figure}

\section{ESTER models} \label{sec:ester}

The study of \cite{Gagnier2019b} presents the rotational evolution of intermediate-mass stars using 2D \ester\footnote{\url{https://github.com/ester-project/ester}} models, where the chemical evolution is emulated with a succession of stationary models with the hydrogen mass fraction in the core given by,
\begin{equation}
    X_{\rm c}(t+\Delta t) = X_{\rm c}(t) - \Delta t \frac{4 m_{\rm p}}{Q}\left< \epsilon \right>_m.
\end{equation}
Here, $m_{\rm p}$ is the proton mass, $Q$ the energy released per reaction, and $\left< \epsilon \right>_m$ the mass averaged nuclear energy production. The models of \cite{Gagnier2019b} do not account for the diffusion of material and thus the composition remains constant throughout the entire radiative envelope. Moreover, they assume a time evolution along a series of quasi-steady states of the star. Here, we use the modified version of \ester~\edit{(r23.09.1-evol)} as presented in \cite{Mombarg2023b} that additionally solves for the chemical diffusion and fully takes into account the time-dependence of structure of the star.
Our physical setup is the same, which we now briefly recapitulate.

Our models include a chemical mixing induced by rotational shear. Indeed, an asteroseismic study of 26 slowly pulsating B-type stars (SPBs) by \cite{Pedersen2021} reveals that, for most stars, SSE models with rotational mixing provide a better solution to the observed gravity-mode pulsation periods, than those with mixing by internal gravity waves. Moreover, a correlation was observed between the efficiency of chemical mixing and the rotation rate. 

In \ester, rotational mixing is based on the work of \cite{Zahn1992} where the vertical diffusion coefficient scales with the shear, $\vn \cdot \nabla \Omega$ (projection perpendicular to the isobars), and the heat diffusivity, $K$, namely,

\begin{equation} \label{eq:Dv}
    D_{\rm v} = \eta \left< N_0^2\right>_V^{-1} \left< K r^2  \left(\vn\cdot\nabla \Omega\right)^2 \right>_\theta.
\end{equation}
Here, $\left< N_0^2\right>_V$ is the volume-averaged value of the Brunt-V\"ais\"al\"a frequency in the radiative envelope and $\eta$ is a free constant that we set to a value such that $\eta \left< N_0^2\right>_V^{-1} = 10^{6}\,{\rm s^2}$. This value prevents numerical issues due to steep chemical gradients building up during evolution. We fix the value of the horizontal chemical diffusion coefficient to $10^5\,{\rm cm^2 s^{-1}}$. 

Furthermore, in \cite{Espinosa2013} and \cite{Gagnier2019b} viscous effects are taken into account in the surface Ekman layers only, while in the new \ester models which we presently use, viscous stresses are included throughout the star. This means we include a constant kinematic viscosity of $\nu = 10^7\,{\rm cm^2 s^{-1}}$ in both the horizontal and vertical direction, when we solve the angular momentum equation,

 \begin{equation}
\dt{s^2\Omega} + \bold{v} \cdot \nabla(s^2 \Omega) = \frac{1}{\rho} \nabla(\rho \nu s^2 \nabla \Omega). \label{eq:omega2}
 \end{equation}
 Here, $s$ is the distance to the rotation axis, $\bold{v}$ is the meridional velocity field, and $\rho$ is the density. We note that this choice for the kinematic viscosity was adequate to reproduce the measured rotation profile of the $\beta$~Cephei pulsator studied by \cite{Burssens2023} at the correct age, as demonstrated in \cite{Mombarg2023b}. As we limit ourselves to masses $\leq 10$\Msun, we assume mass loss can be neglected. 

\begin{figure*}[htb]
    \centering
    \includegraphics[width = 0.9\textwidth]{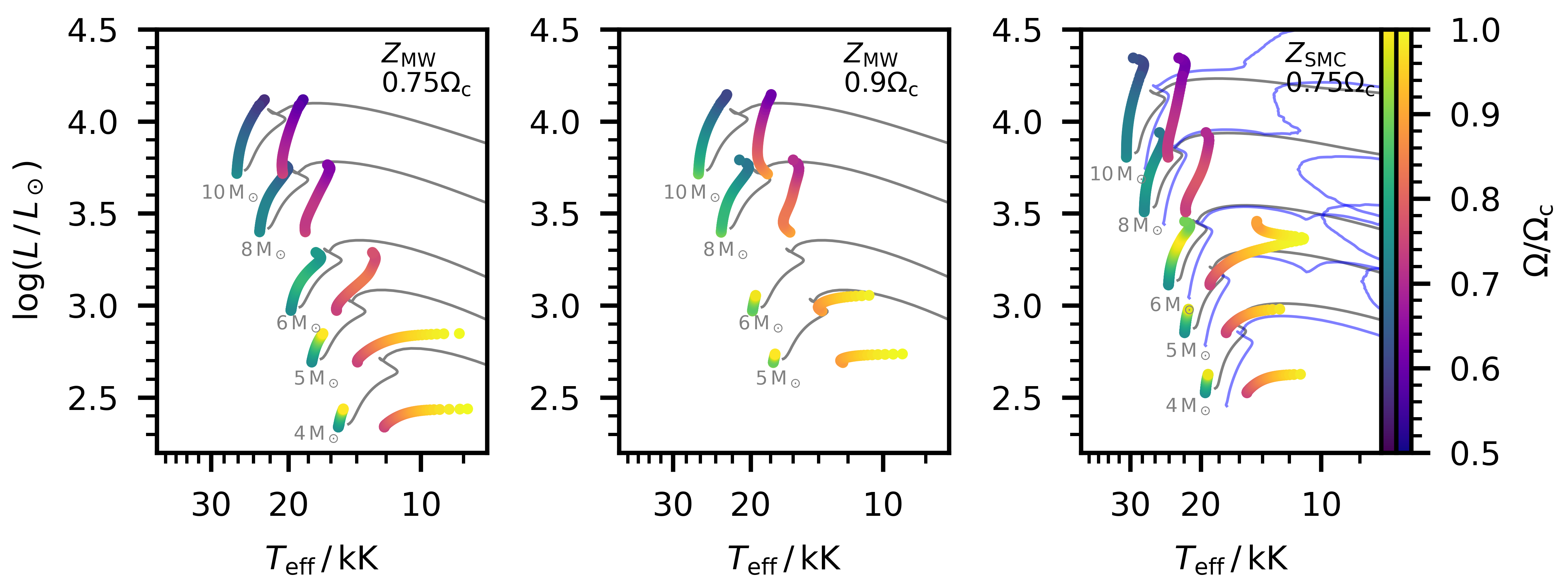}
    \caption{Hertzsprung-Russell diagram showing the evolution of the \ester models, colour-coded by the fraction of critical rotation. For each mass, a track is plotted once using the effective temperature at the pole and once at using the effective temperature the equator. The left (right) colour bar corresponds to the pole (equator). For comparison, non-rotating \mesa (r23.05.1; \citet{Paxton2011, Paxton2013, Paxton2015, Paxton2018, Paxton2019, Jermyn2023}) tracks are shown with the grey lines. The \mesa tracks were computed up to the base of the red giant branch, whereas the \ester tracks are stopped close to the TAMS or when critical rotation is reached. \edit{In the right panel, also rotating 1D models (same initial rotation) from \cite{Georgy2013} are shown in blue.} }
    \label{fig:HRD}
\end{figure*}

The \ester models are discretized in 12 domains, which are spheroidal shells separated by isobars \citep{Rieutord2016}. Figure~\ref{fig:AM_distr} shows the distribution of angular momentum per domain for a 10\Msun star. From this figure, we can conclude that most of the star's angular momentum is located in the deep interior and therefore any angular momentum extracted from the stellar surface via winds is extremely small in the mass regime we study here. As we do not assume solid-body rotation, the angular momentum loss at the surface is less than predictions from 1D models like the ones used by \cite{Hastings2020}. In Appendix~\ref{ap:omega} we show an example of a 2D rotation map at the zero-age main sequence (ZAMS), as well as for a model at the terminal-age main sequence (TAMS), and a model rotating at the critical rotation rate. The Keplerian critical angular velocity that we use in this paper as the point of criticality is defined by

\begin{equation}
    \Omega_{\rm c} = \sqrt{\frac{G M_\star}{R_{\rm eq}^3}},
\end{equation}
where $R_{\rm eq}$ is the radius at the equator. \cite{Gagnier2019a} have shown that the reduction of $\Omega_{\rm c}$ due to the radiative acceleration at the stellar surface can be neglected in the mass regime we study here. 

We investigate the rotational evolution for both a metallicity typical for stars in the Milky Way (MW, $Z_{\rm MW} = 0.02$), and for a metallicity typical for stars in the Small Magellanic Cloud (SMC, $Z_{\rm SMC} = 0.003$). Lastly, we start our models at the ZAMS with a certain fraction of the critical angular velocity. These initial models are in a steady state like those of \cite{Espinosa2013}. Throughout this paper, the mentioned ages concern the time spent on the MS. Pre-MS lifetimes range from 4.4\,Myr for 4\Msun to 0.34\,Myr for 10\Msun. 

\section{Evolution of rotation} \label{sec:rot_evol_MW}

We have computed evolution tracks for masses 4, 5, 6, 8, and 10\Msun, all starting with an initial rotation rate of 75\percent of the critical rotation frequency\footnote{\edit{All models presented in this paper are available on \url{doi.org/10.5281/zenodo.10411194}.}}. 

As we do not take into account any physics for the formation of a circumstellar disk, we stop the models once critical rotation is reached. Table~\ref{tab:veq} lists the corresponding initial velocities at the equator. 

\subsection{Model predictions}
The evolution tracks in the HRD are shown in Fig.~\ref{fig:HRD} for the effective temperature at the pole and at the equator. It should be noted that for stars close to the critical rotation limit, the difference between the effective temperature at the pole and equator can differ by as much as a factor two. \edit{The hooks seen at the end of some evolution tracks are the result of core contraction due to the steep drop in hydrogen in the convective core. Some models were stopped before the hook due to numerical problems. The hook seen in the equatorial-\teff track for the 6\Msun model in the right panel of Fig.~\ref{fig:HRD} is the result of the star reaching near critical rotation for a brief moment, but then decreasing the fraction of critical rotation again. For the $Z_{\rm SMC}$-case, we also show the 1D rotating evolution tracks from \cite{Georgy2013} (not available for $Z_{\rm MW} = 0.02$ as we assume in this paper). In general, the \teff predicted from the 1D rotating models is the average of the polar and equatorial \teff predicted by our \ester models. The 1D models remain below $\Omega/\Omega_{\rm c} = 0.8$, while our 2D models predict stars below 8\Msun to reach critical rotation during the main sequence.  }

\begin{figure}[htb]
    \centering
    \includegraphics[width = 0.65\columnwidth]{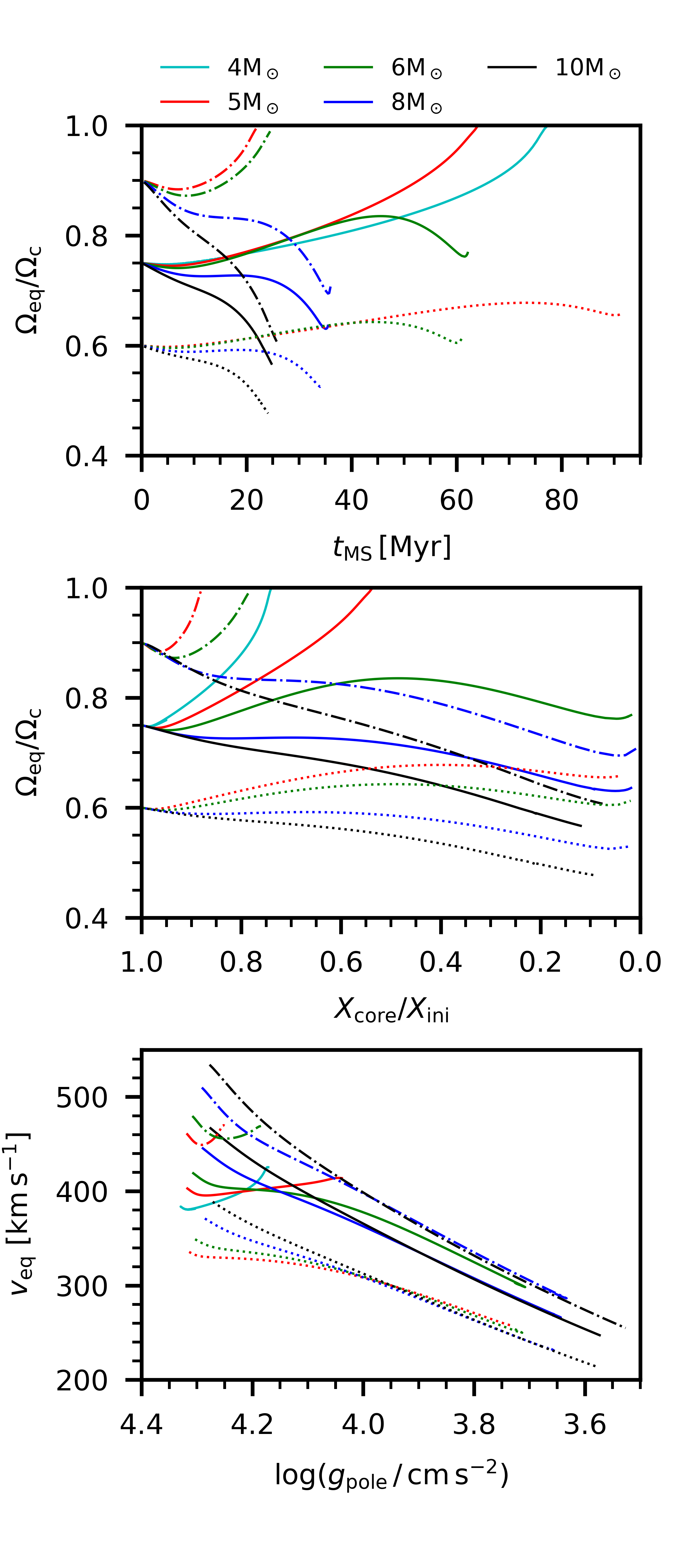}%0.85
    \caption{Evolution of the fraction of critical rotation as a function of time spent on the MS (top panel) and as a function of the mass fraction of hydrogen in the core with respect to the initial mass fraction (middle panel). The bottom panel shows the velocity at the equator as a function of the surface gravity at the pole. The different line styles correspond to the different initial rotation rates at the start of the MS. The models shown here are for $Z_{\rm MW}$.}
    \label{fig:Omega_evol_MW}
\end{figure}

Figure~\ref{fig:Omega_evol_MW} shows the evolution of the fraction of critical rotation \omegac as a function of time spent on the main sequence and as a function of the hydrogen mass fraction in the core ($X_{\rm c}$).  We note several interesting findings. We find that for \omegai$ = 0.75$, the stars of 4, 5 and 6\Msun increase their fraction of critical rotation during the MS. While the 6\Msun star spins down\footnote{In this paper, the term "spinning down/up" refers to the evolution of the fraction of critical angular velocity, not to the real angular velocity. The actual angular velocity decreases as the star evolves, but the critical angular velocity may decrease even faster moving the star closer to critical angular velocity. We call this motion towards criticality ``spin-up''.} after about 45\,Myr when it reaches a maximum of \omegac = 0.84, the 4 and 5\Msun models continue to spin up and reach critical rotation around 77\,Myr and 64\,Myr, respectively. Our models for 8 and 10\Msun predict a continuous decrease in the fraction of critical rotation during the MS. 

Additionally, we have run models starting at 90\percent of the initial critical rotation frequency. At this very high initial rotation rate, the 5\Msun star reaches critical rotation after roughly 22\,Myr, and the 6\Msun star now also spins up to critical rotation only a few Myr after. In case of a 8\Msun star, even an initial rotation rate as high as \omegai = 0.9 is not sufficient to reach critical rotation during the MS, and the star spins down and reaches \omegac = 0.7 near the end. Finally, we have also tested whether an initial rotation rate of 60\percent of the critical rotation is still enough to spin up a 5\Msun star to criticality. As can been seen in Fig.~\ref{fig:Omega_evol_MW} (red dotted line), the star reaches a maximum of roughly \omegac = 0.7 and afterwards spins down. In summary, the threshold on the minimum initial rotation rate needed to reach critical rotation during the MS decreases with decreasing mass. 

\subsection{Uncertainties from chemical mixing}

As mentioned before, the feedback of a chemical gradient on the efficiency of chemical diffusion is not taken into account in the \ester models and Eq.~(\ref{eq:Dv}) has a free parameter. In \cite{Mombarg2023b}, we have shown that this description of chemical mixing does reproduce the measured core mass by \cite{Burssens2023}. The choice for the value of $\eta$ will influence the exact age at which the star reaches critical rotation. More efficient mixing will result in a larger convective core and a slightly larger radiative envelope, thereby increasing the time scale on which baroclinic modes are damped and angular momentum is redistributed \citep[Eq.~(3) in ][]{Mombarg2023b}. We have recomputed the evolution track of the (5\Msun, \omegai = 0.75, $Z_{\rm MW}$) model, where we have changed $\eta = 1.85$ to 1. Such a change in the global scaling of the vertical diffusion coefficient increases the time to reach criticality to 67\,Myr, that is, 3\,Myr later (see Appendix~\ref{ap:mixing}). Therefore, our conclusions will remain the same for changes to the value of $\eta$ that are of order unity. \newline

\subsection{Comparision with observations}

\begin{figure}[htb]
    \centering
    \includegraphics[width = 0.95\columnwidth]{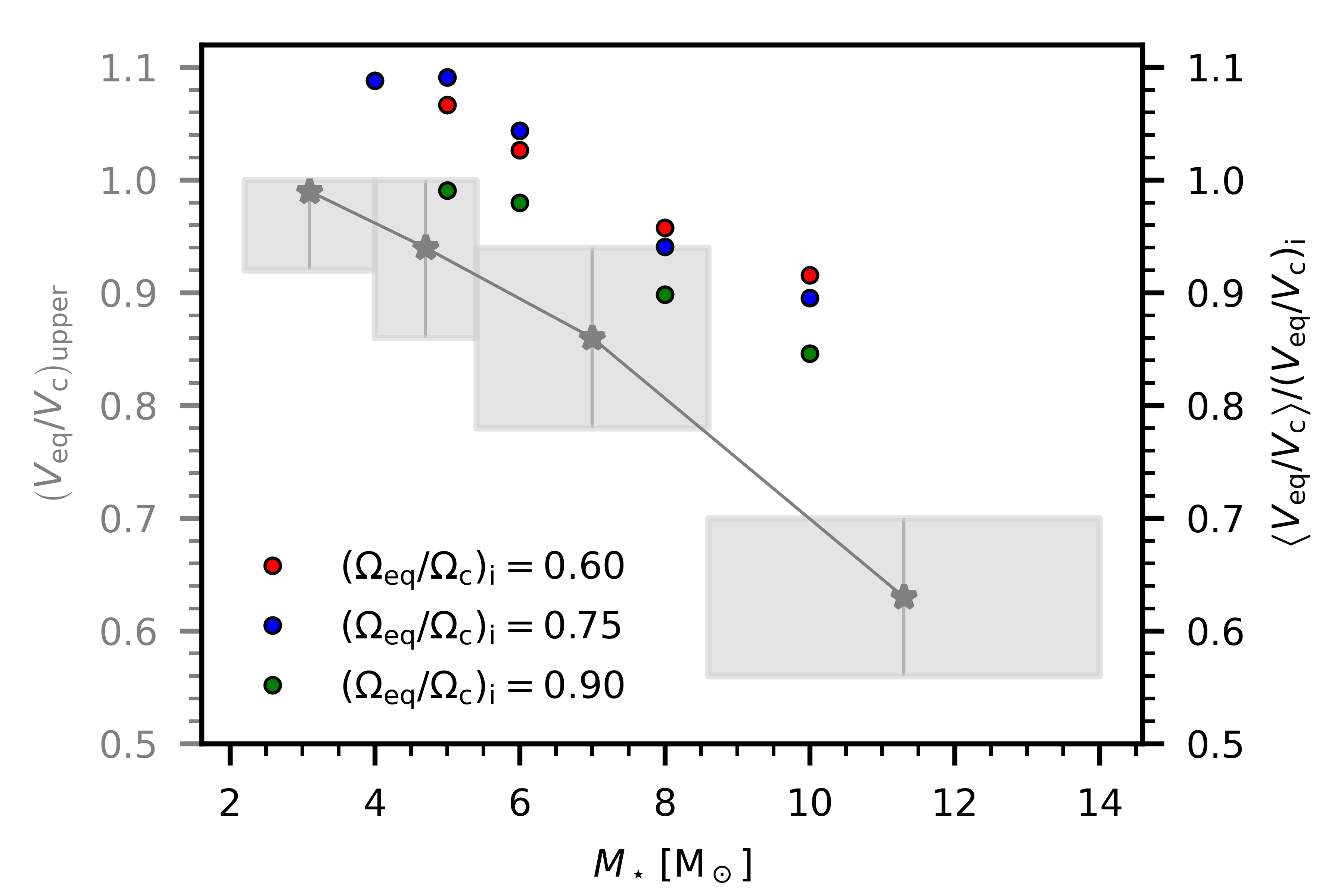}
    \caption{The average value of the equatorial velocity with respect to the critical one, $\left< {\rm V}_{\rm eq}/{\rm V}_{\rm c}\right>$, divided by the initial value at the ZAMS, as a function of stellar mass. Data points in grey show the observationally derived upper limits of ${\rm V}_{\rm eq}/{\rm V}_{\rm c}$, taken from the last column of Table 7 of \citet[][]{Huang2010}. }
    \label{fig:logg_veq}
\end{figure}

The work of \cite{Huang2010} presents the ${\rm V} \sin i / {\rm V}_{\rm c}$ distribution\footnote{The critical rotation velocity is estimated from the Roche model.} of a large sample of cluster and field B-type stars according to the spectroscopic surface gravity at the rotational pole, $\log g_{\rm pole}$ (Be stars excluded). These authors find that stars between 2 and 4\Msun on average have larger values for ${\rm V} \sin i / {\rm V}_{\rm c}$ than stars between 7 and 13\Msun. In Fig.~\ref{fig:logg_veq}, we show for each mass ${\rm V}_{\rm eq}/{\rm V}_{\rm c}$ averaged over the MS life time (or the time until critical rotation is reached) computed from the 2D \ester models for $Z_{\rm MW}$. The points are normalised to the initial value at the ZAMS, $({\rm V}_{\rm eq}/{\rm V}_{\rm c})_{\rm i}$. We observe a decrease in $\left< {\rm V}_{\rm eq}/{\rm V}_{\rm c}\right>/({\rm V}_{\rm eq}/{\rm V}_{\rm c})_{\rm i}$ with respect to the stellar mass that has a similar slope compared to upper limits of ${\rm V}_{\rm eq}/{\rm V}_{\rm c}$ derived by \cite{Huang2010} for non-emission B-type stars (grey points in Fig.~\ref{fig:logg_veq}). In other words, observations show that the upper limit on ${\rm V}_{\rm eq}/{\rm V}_{\rm c}$ decreases with mass, while our models show the same trend once ${\rm V}_{\rm eq}/{\rm V}_{\rm c}$ is rescaled by its initial value at the ZAMS. This latter rescaling could be avoided by running a large set of models representing the evolution of a star cluster and computing the maximum angular velocity per bin of mass, but such a simulation is presently too demanding in computing time.

\section{Effect of metallicity} \label{sec:rot_evol_SMC}

\begin{figure}
    \centering
    \includegraphics[width = 0.85\columnwidth]{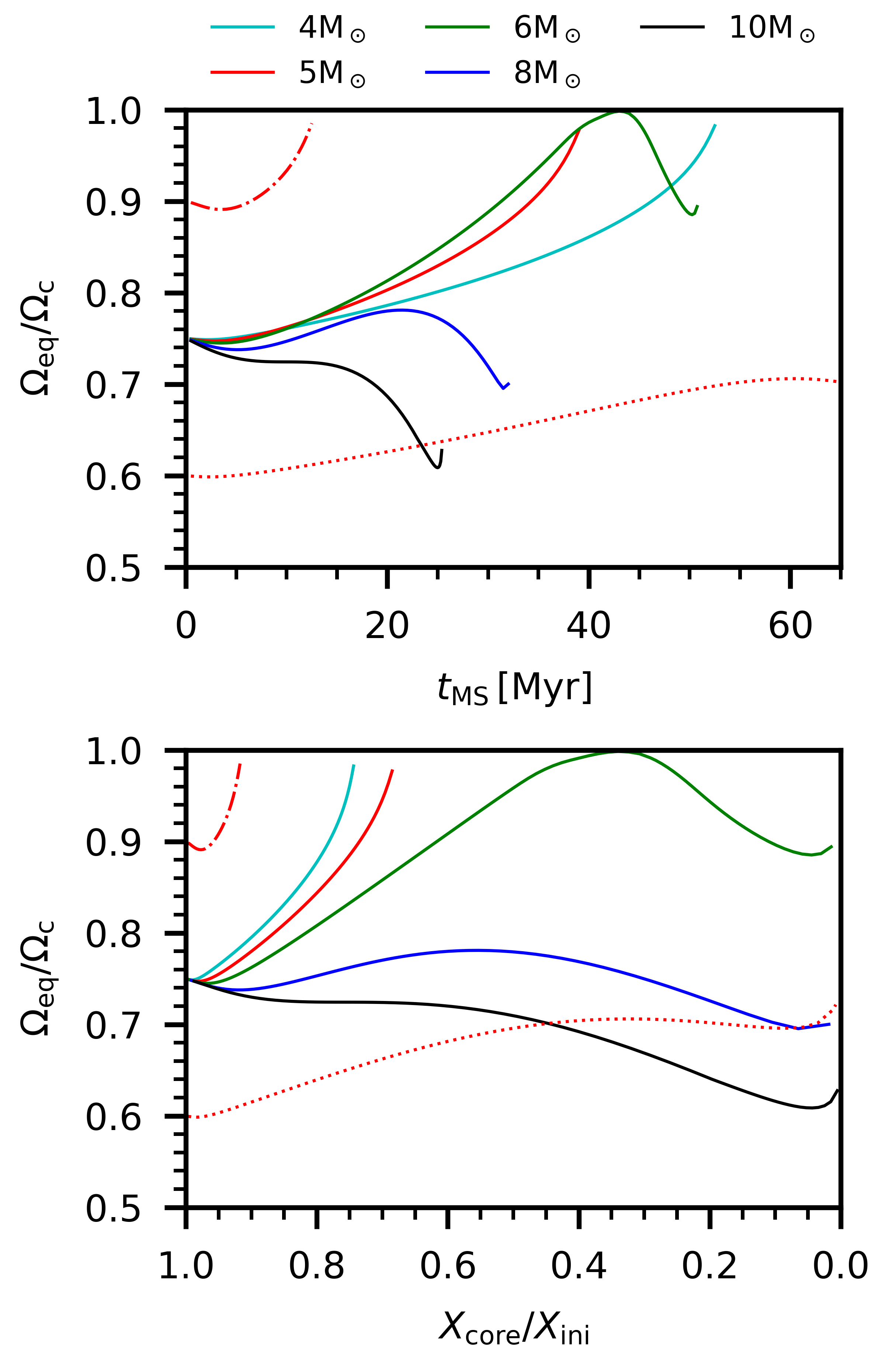}
    \caption{Same as Fig.~\ref{fig:Omega_evol_MW}, but for $Z_{\rm SMC}$.}
    \label{fig:Omega_evol_SMC}
\end{figure}

We have also computed evolution tracks for the same masses with a metallicity typical of the SMC. The results are shown in Fig.~\ref{fig:Omega_evol_SMC}. We find that at lower metallicities, the stars spin up faster, and the 5\Msun star reaches critical rotation around 40\,Myr, whereas for Galactic metallicity, this would take 63\,Myr. Similarly, the 5\Msun star starting with \omegai = 0.9 reaches critical rotation after 13\,Myr at SMC metallicity, whereas it takes 22\,Myr at MW metallicity. Moreover, the lower metallicity is sufficient to also spin up the 6\Msun star to (near) critical rotation around 45\,Myr.
Since the time scale on which baroclinic modes are damped scales with the square of the thickness of the radiative envelope \citep{busse81}, we expect angular momentum to be faster redistributed in lower-metallicity stars as they are more compact. On the other hand, the nuclear time scale will also decrease with decreasing metallicity since a lower metallicity results in higher central temperatures. Nonetheless, we find that for the 5\Msun ZAMS model, the baroclinic time scale differs with 150\percent between $Z_{\rm MW}$ and $Z_{\rm SMC}$, while the nuclear time scale only differs by 25\percent. Likewise, for the 10\Msun ZAMS model, the difference in the baroclinic time scale is 50\percent, compared to 14\percent for the nuclear time scale. Again, our models predict no stars $\geq 8$\Msun that significantly increase the fraction of critical rotation compared to their initial value when \omegai equals 0.75. Also, even at SMC metallicity the spin up of the 5\Msun star starting with \omegai = 0.6 stalls after 60\,Myr having reached a maximum of \omegac = 0.7.

It is interesting to note that our 2D models suggest an increase in the fraction of Be stars with decreasing metallicity. Indeed, larger fractions of Be stars have been observed in lower metallicity environments \citep{Martayan2010, Iqbal2013}, but this is not reproduced by the 1D models by \cite{Ekstrom2008, Ekstrom2012} and \cite{Granada2013}. The predictions of \citet[][also based on 1D models]{Hastings2020} do predict this observed relation between the fraction of Be stars and the metallicity, but in their models, this is due the dependence of the mass-loss rate on the metallicity. Thus, this should only be of importance in more massive stars, namely above 10\Msun, while we find a strong metallicity effect also at lower masses and we do not account for mass loss.  

\begin{table}[]
    \centering
    \caption{The equatorial velocity at the ZAMS. }

    \begin{tabular}{ccccc}
    \hline \hline
    $M_\star/{\rm M_\odot}$  & $Z$ & \omegai & $v_{\rm eq}\,[{\rm km\,s^{-1}}]$ & $t_{\rm c}\,[{\rm Myr}]$ \\
    \hline
        \hline
        5 & 0.02 & 0.60 & 336 & - \\
        6 & 0.02 & 0.60 & 350 & - \\
        8 & 0.02 & 0.60 & 372 & - \\
        10 & 0.02 & 0.60 & 391 & - \\
        4 & 0.02 & 0.75 & 384 & 77 \\
        5 & 0.02 & 0.75 & 403 & 64 \\
        6 & 0.02 & 0.75 & 420 & - \\
        8 & 0.02 & 0.75 & 447 & - \\
        10 & 0.02 & 0.75 & 469 & - \\
        5 & 0.02 & 0.90 & 462 & 22 \\
        6 & 0.02 & 0.90 & 481 & 25\\
        8 & 0.02 & 0.90 & 512 & - \\
        10 & 0.02 & 0.90 & 537 & - \\        
        \hline
        5 & 0.003 & 0.60 & 380 & -\\
        4 & 0.003 & 0.75 & 437 & 53 \\
        5 & 0.003 & 0.75 & 457 & 40 \\
        6 & 0.003 & 0.75 & 473 & 43 \\
        8 & 0.003 & 0.75 & 501 & - \\
        10 & 0.003 & 0.75 & 523 & - \\
        5 & 0.003 & 0.90 & 523 & 13\\
        
    \hline  

    \end{tabular}
    \tablefoot{The last column indicates the time spent on the MS before reaching critical rotation (no value means critical rotation is not reached).}
    \label{tab:veq}
\end{table}

\section{Conclusions} \label{sec:conclusions}

Following the successful calibration of the first 2D stellar structure and evolution models by \cite{Mombarg2023b}, 
we have presented here rotating stellar evolution tracks for intermediate-mass stars computed with the 2D \ester code. Solving the stellar structure in two dimensions instead of one has the added value that we account for the centrifugal deformation of the star and we compute the rotation profile self-consistently from the baroclinic torque (e.g. Eq.~(26) in \citealt{Espinosa2013}). We have presented evolution models for stellar masses 4, 5, 6, 8 and 10\Msun for initial rotation rates between 60 and 90\percent of the critical angular velocity. Summarising, the most important findings from our 2D models are the following.

\begin{itemize}
    \item An initial rotation rate of \omegai = 0.75 is not enough for stars with mass $\geq 6$\Msun and galactic metallicity to evolve towards critical rotation. Stars of 5\Msun will reach critical rotation around 60\,Myrs.
    
    \item At an initial rotation rate of \omegai = 0.9, stars of 5 and 6\Msun do reach critical rotation around 20-25\,Myrs. 
    
    \item Stars in the SMC, thus with a lower metallicity than Galactic stars, spin up faster and stars with masses 5 and 6\Msun\ can reach critical rotation around 40\,Myr.
    \item Stars with mass $\geq 8$\Msun\ always move away from critical rotation during the main sequence, even for SMC metallicity or \omegai = 0.9. 
\end{itemize}

If we assume that the emergence of the Be phenomenon requires the star to rotate close to criticality, we can make the following predictions for the fraction of single Be stars with respect to the total number of B-type stars. First, if we hypothesise that single Be stars are born as rapid rotators, then our models show that the minimum initial rotation rate needed to spin up stars to criticality decreases with decreasing metallicity. Second, our models predict that no stars with mass above 8\Msun\ can stay rotating at or near critical rotation. We thus conclude that for massive B-type stars $\gtrsim$~8\Msun the velocity threshold to expel matter from the surface must be lower than for the lower-mass ones. 

\edit{In order to constrain the prevalence of the single-star formation channel compared to the binary formation channel, the upper limit on the fraction of critical rotation at which material can be expelled from the surface (for a given stellar mass) needs to be quantified. Additionally, while the predicted rotation profile from \ester are in agreement with the observed one in case of the $\beta$~Cephei pulsator HD192575 \citep{Mombarg2023b}, future measurements of rotation profiles of massive stars with a precise mass and age will help us get a better picture of the accuracy of the current treatment of angular momentum transport.  }

\edit{Even though we have limited ourselves to masses up to 10\Msun, we expect the more massive B-type stars that weigh up to 16\Msun to also decrease the fraction of critical rotation during the MS. Assuming a mass loss rate predicted by \cite{Bjorklund2021}, a 16\Msun has lost about 0.02\Msun by the point of core-hydrogen exhaustion (or about a factor 10 more when assuming \cite{Vink2001} mass loss rates). Hence, the decrease in the value of $\Omega_{\rm c}$ due to mass loss is negligible. Moreover, \cite{Gagnier2019a} have demonstrated that the reduction of the critical rotation velocity due to radiative acceleration is small for stars below 40\Msun. However, the decrease of \omegac with respect to time can be accelerated as a result of angular momentum loss at the stellar surface \citep{Gagnier2019b}. While we expect this effect to be small, 2D evolution with mass loss need to be further studied to quantify this.   }

Observed fractions of Be stars compared to the total number of B-type stars in clusters show little increase over time \citep{McSwain2005}. In light of our results for single stars, this would imply that the majority of stars is born with a rotation rate $\lesssim$60\percent of the critical angular velocity at the ZAMS. The higher fraction of Be stars in low-metallicity environments observed around an age of 40\,Myr \citep{Martayan2010} is in line with our predictions as we predict a lower threshold on the initial rotation for stars to reach critical rotation and a smaller time span to do so. Furthermore, the predicted evolution of the surface velocity by the \ester models seems to be in good agreement with the measurements of the fastest rotating B-type stars in the sample of \cite{Huang2010}.

In summary, the results of two-dimensional stellar structure and evolution models that include differential rotation and meridional currents show the limitations of one-dimensional models to predict the evolution of rapidly rotating stars. In addition to the (projected) velocity at the surface, the surface abundance of the nitrogen-14 isotope is also a useful observable to test the theory of rotational evolution in massive stars. In stars that are undergoing hydrogen burning via the CNO-cycle, an overabundance of nitrogen is created in the core and the enhancement of its abundance at the surface is indicative of the efficiency of chemical (rotational) mixing \citep[e.g.][]{Hunter2009,Brott2011}. In a future paper, we will investigate the evolution of surface abundances predicted from rotational mixing with 2D evolution models.

%\clearpage
\begin{acknowledgements}
 The research leading to these results has received funding the French Agence Nationale de la Recherche (ANR), under grant MASSIF (ANR-21-CE31-0018-02). MR also acknowledges the support from the Centre National d'Etudes Spatiales (CNES) and  from the European Research Council (ERC) under the Horizon Europe programme (Synergy Grant agreement N$^\circ$101071505: 4D-STAR).  
Computations of \ester 2D-models have been possible thanks to HPC resources from CALMIP
supercomputing center (Grant 2023-P0107). While partially funded by the European Union, views and opinions expressed are however those of the authors only and do not necessarily reflect those of the European Union or the European Research Council. Neither the European Union nor the granting authority can be held responsible for them. The authors are thankful for the feedback on the manuscript provided by the anonymous referee. 
\end{acknowledgements}

\bibliographystyle{aa} % style aa.bst
\bibliography{main} % your references Yourfile.bib

\clearpage

\appendix 
\section{2D rotation maps} \label{ap:omega}
This appendix shows the 2D maps of the angular rotation velocity for a 5\Msun model at the ZAMS, a 5\Msun model when it reaches critical rotation, and a 6\Msun model near TAMS. The rotation map of the 6\Msun model at the ZAMS (not shown) is very similar to that of the 5\Msun model.
\begin{figure}[htb]
    \centering
    \includegraphics[width = 0.6\columnwidth]{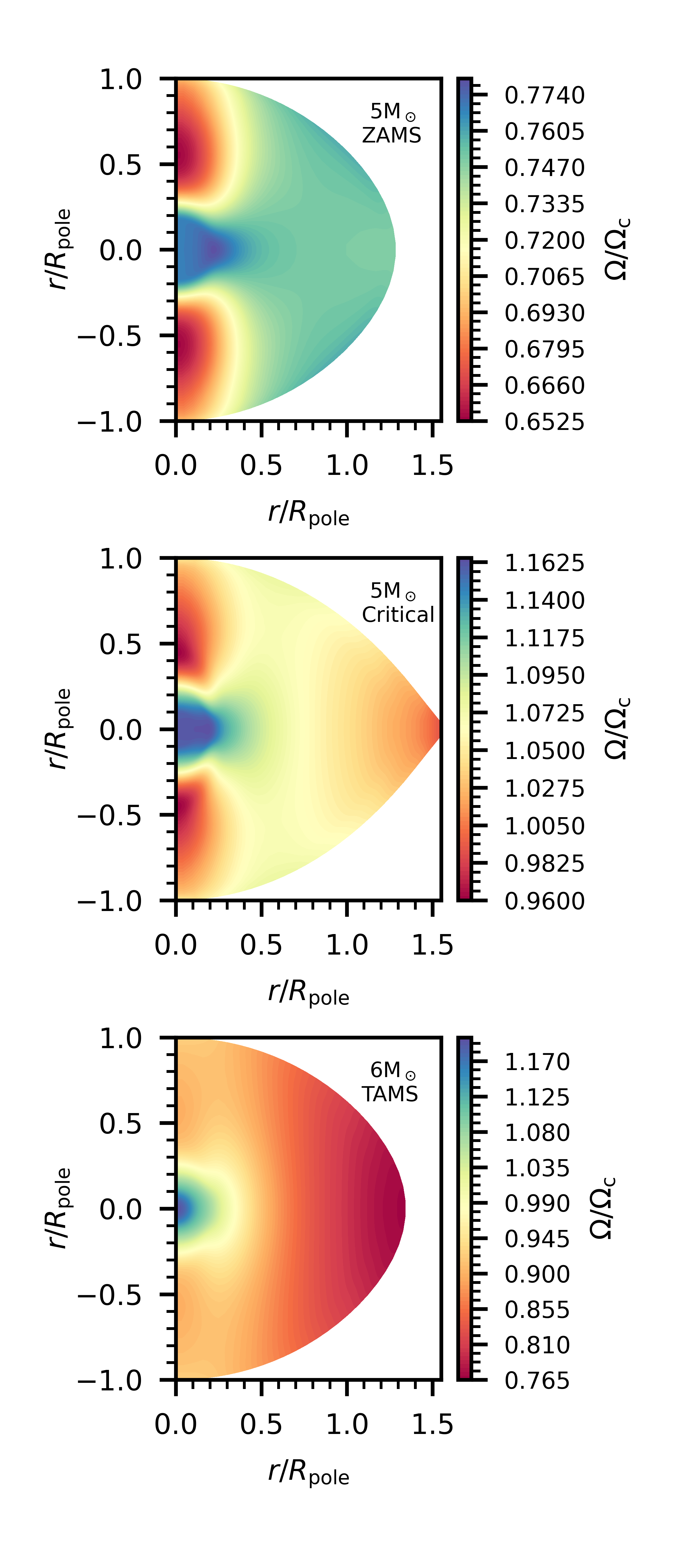}%0.9
    \caption{Two-dimensional maps of the angular rotation velocity with respect to the Keplerian critical rotation velocity.}
    \label{fig:Omega_2d}
\end{figure}

\section{Influence of chemical mixing} \label{ap:mixing}

This appendix shows the influence of the chemical mixing parameter $\eta$ on the rotational evolution.
\begin{figure}[htb]
    \centering
    \includegraphics[width = 0.9\columnwidth]{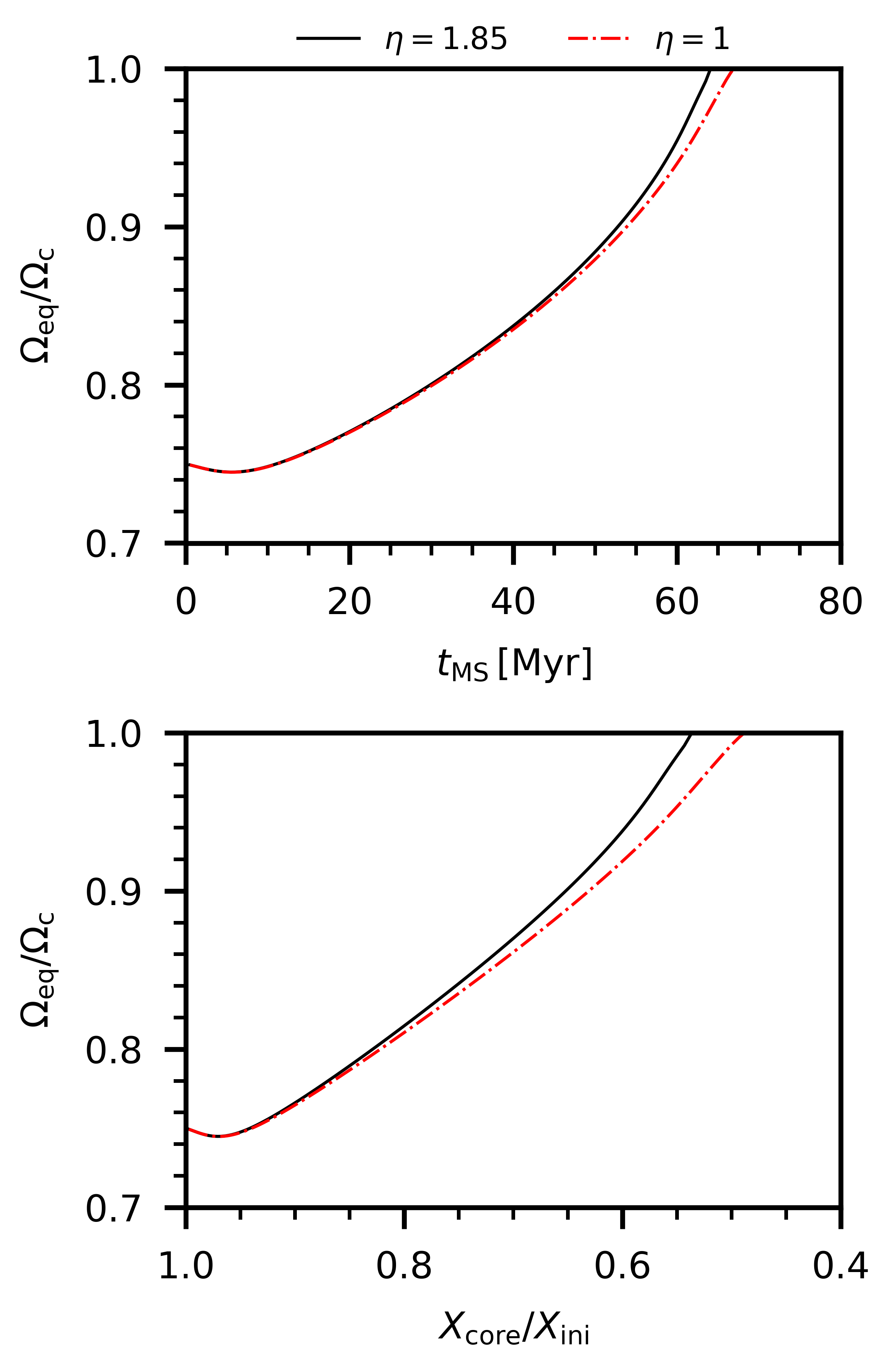}
    \caption{Evolution of the fraction of critical rotation as a function of time spent on the MS (top panel) and as a function of the mass fraction of hydrogen in the core with respect to the initial mass fraction. The different line styles correspond to different global scalings of the vertical diffusion coefficient predicted from rotational mixing (Eq.~\ref{eq:Dv}). the models are shown for a 5\Msun star with $Z_{\rm MW}$.}
    \label{fig:Omega_evol_mix}
\end{figure}


\end{document}